\documentclass[11pt,aps,superscriptaddress,notitlepage,pra]{revtex4-1}%
\usepackage{amsmath}
\usepackage{amssymb}
\usepackage{amsfonts}
\usepackage{revsymb}
\usepackage{graphicx}
\usepackage{color}
\usepackage{hyperref}%
\usepackage{graphicx}
\usepackage{tabularx}
\usepackage{pgfplots}
\usepackage{siunitx}
\usepackage{upgreek}
\providecommand{\U}[1]{\protect\rule{.1in}{.1in}}
\newtheorem{lemma}{Lemma}[section]

\newtheorem{proposition}[lemma]{Proposition}

\begin{document}

\author{Martin Marek}
\affiliation{GJH, Joint School Novohradsk\'a, Bratislava, Slovakia}
\author{Matej Badin}
\affiliation{FMFI, Comenius University, Bratislava, Slovakia}
\author{Martin Plesch\thanks{martin.plesch@savba.sk}}
\affiliation{Institute of Physics, Slovak Academy of Sciences, Bratislava, Slovakia}
\affiliation{Institute of Computer Science, Masaryk University, Brno, Czech Republic}

\title{Gee-Haw Whammy Diddle}

\begin{abstract}
Gee-Haw Whammy Diddle is a seemingly simple mechanical toy consisting of a
wooden stick and a second stick that is made up of a series of notches with a
propeller at its end. When the wooden stick is pulled over the notches, the
propeller starts to rotate. In spite of its simplicity, physical principles
governing the motion of the stick and the propeller are rather complicated and
interesting. Here we provide a thorough analysis of the system and parameters
influencing the motion. We show that contrary to the results published on this
topic so far, neither elliptic motion of the stick nor frequency
synchronization is needed for starting the motion of the propeller.

\end{abstract}
\maketitle

\section{Introduction}

Scientific toys play a very important role in science education and
popularization of science. Often they are used for demonstration purposes
connected with a specific scientistic phenomenon or principle. Students or
general audience can get a hands-on feeling of eddy currents by throwing a
magnet into a metallic tube and observing its slow motion, or a better
understanding of magnetic field by levitating a magnetic spinner.

\begin{figure}
\begin{center}
\includegraphics[width = 0.5\linewidth]%
{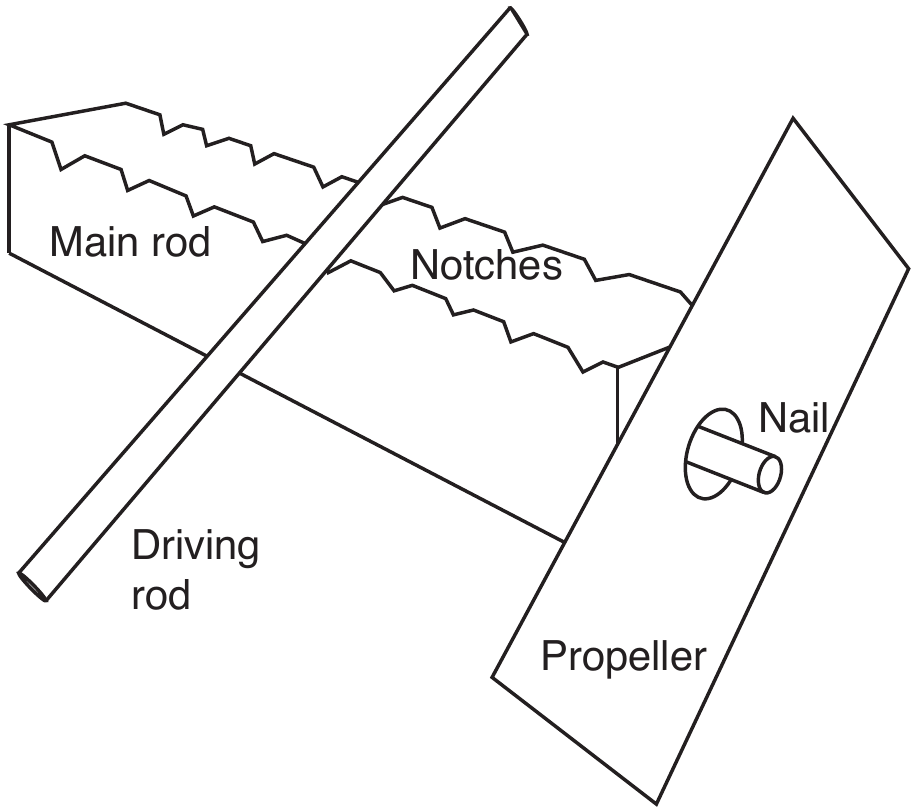}%
\caption{A typical gee-haw whammy diddle}%
\label{Gee_haw}%
\end{center}
\end{figure}

Gee-haw whammy diddle is a scientific toy of a slightly different type. It
consist of a wooden base-stick equipped with a series of notches on one of its
sides with a propeller loosely attached to its end so that it can freely
rotate (Fig. \ref{Gee_haw}).

Another stick is used to rub the first one on the notches,
which causes the propeller to rotate. Skilled performers are able to achieve
very high rotation speed of the propeller and change the rotation direction
very quickly.

Being it a mechanical device, one would expect to be able to pinpoint a simple
and clear explanation, why and how a linear motion along the notches causes a
circular motion of the propeller and what defines the direction of this
motion. This is however not the case in spite of the fact that research on
this topic lasts for almost a century. In one of the first works on this topic
Leonard \cite{Leonard} suggests that the propeller vibrates due to elliptical
motion of the end of the stick and that the direction of the motion is
determined by the phase shift between the vertical and horizontal stick
vibrations, caused by different rubbing directions.

This idea was further developed in \cite{Schichting} where the direction of
the rotation is suggested to be connected to the way the stick is being held
by hand. Authors also provide basic numerical outputs given the possibilities
of computers of that period. Almost a decade later three other papers appeared
on this topic. M. J. Clifford et. al. \cite{Clifford} suggested to link the
problem to chaotic orbits of parametrically excited pendulum. Jun Satonobu et.
al. \cite{Satonobu} provided probably the most complex analysis of the problem
from both theoretical and experimental point of view so far and concluded the
validity of the original hypothesis that the rotation direction is determined
by the elliptical movement of the end of the stick and its areal velocity.
However the presented experiment concluded only one-side implication of the
hypothesis, namely that an elliptical movement of the stick causes the
rotation of the propeller into given direction, but did not show that this is
the only/prevailing cause of the movement in the real toy.

Further analytical and experimental research was performed by Wilson in
\cite{Wilson}. Results presented here partially contradict the previous work
in claiming that the direction of the rotation of the end of the stick does
not necessarily defines the direction of the propellers motion. He also
presented the hypothesis that the frequency of rotation of the propeller is
synchronized with the driving frequency of the notches. However this work has
a few limiting factors incorporated. In the analytical part no slip is assumed
between the stick and the propeller, which makes the problem and its solution
similar the hula-hoop problem, however there is no experimental evidence for
this assumption. In the experiment, recordings were performed with very low
time resolution (30 fps) which apparently led to rather imprecise results of
the motion of the stick. Recently in \cite{Bhattacharjee} the problem was
re-considered by connecting it to Kapitza pendulum and recalling the idea of
synchronization between the motion of the stick and the propeller.

In this paper we perform full experimental analysis of the gee-haw whammy
diddle toy. We concentrate on the way how different types of movements of the
end of the stick cause the propeller to rotate and analyze not only the
movement direction, but also its speed depending on various parameters of the
movement (amplitude, frequency, areal velocity etc.). We show that most of the
hypothesis presented in the literature so far fail under rigorous experimental
tests - areal velocity (elliptical movement) of the stick is not needed to
excite the propeller to high speed rotations and this speed is not connected
to the frequency of vibrations. We suggest that the rotation is a result of a
simple positive-feedback mechanism between the motion of the stick and the
propeller and its direction (in the absence of the areal velocity) is given by
a random fluctuation. The speed of rotation is determined by a combination of
amplitude and frequency of the stick vibration and the highest speeds are
achieved when the typical stick acceleration is about the gravitational acceleration.

\section{Experiment}

Gee-haw whammy diddle in its original form is practically an uncontrollable
device. The fact that it is hold in one hand and rubbed by the other one does
not allow any reasonable control of the excitation parameters, neither the
exact position of the stick, nor its movement. In previous works
\cite{Satonobu, Wilson} different mechanical devices were suggested that
partially fixed the issue of holding the stick, but kept the rather
complicated way of stick excitation by rubbing. Although this mimics the
original idea of the device, it is connected with many complications. First,
as the rubbing stick needs to moved there and back in a periodic movement, the
frequency of the oscillations (caused by running through the notches) if far
from being constant. The amplitude of oscillations also depends on the
position of the stick (notches closer to the end of the stick cause smaller
amplitude then the ones closer the the fixing) and on its speed (higher speed
means stronger collisions). By keeping this system in its original form it is
nearly impossible to separate relevant parameters that influence the motion of
the propeller.

\begin{figure}
\begin{center}
\includegraphics[width = 0.7\linewidth]%
{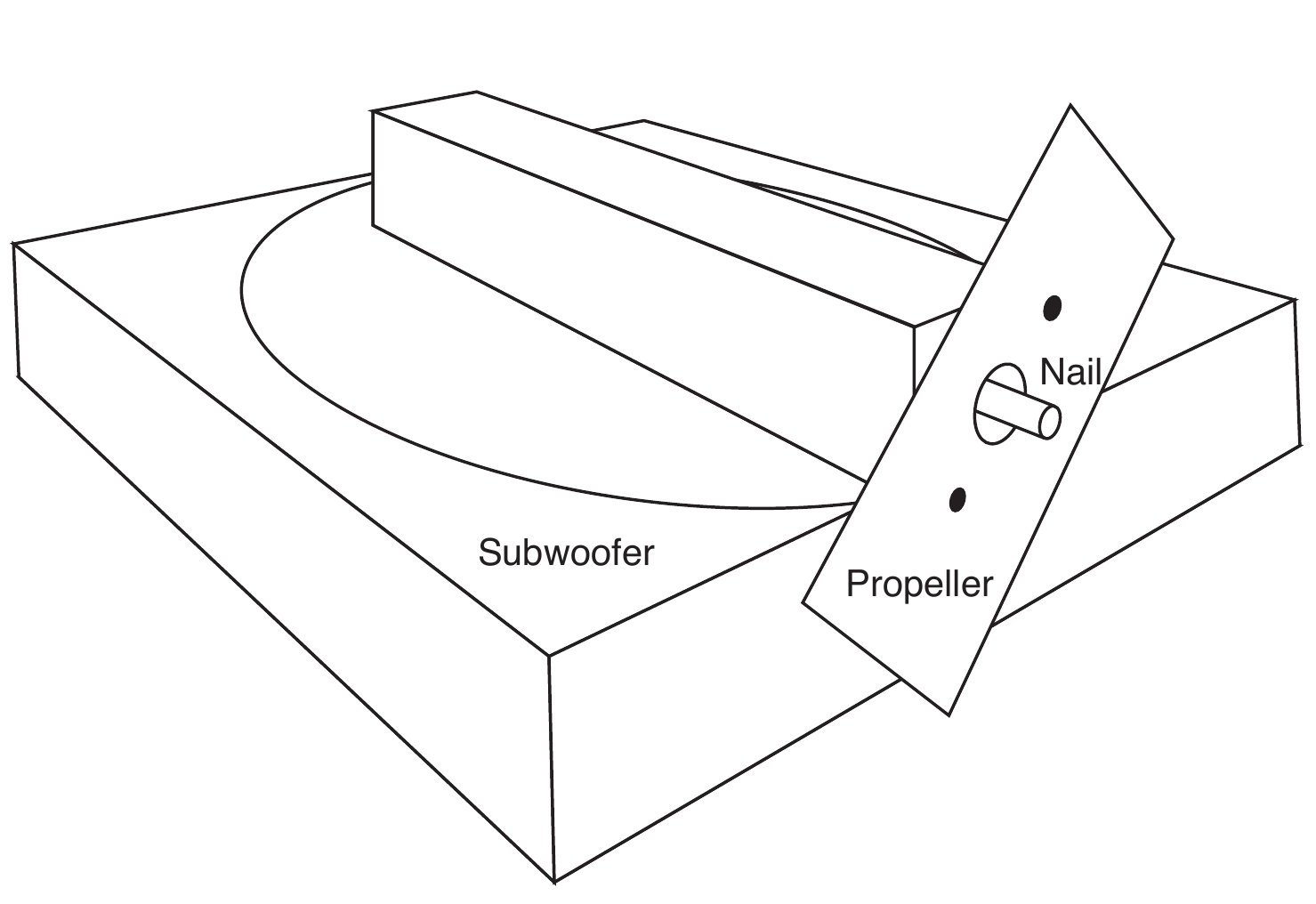}%
\caption{A rod attached directly to a subwoofer.}%
\label{apparatus}
\end{center}
\end{figure}

Therefore we decided to redesign the experiment to concentrate solely on the
motion of the end of the stick. Instead of rubbing it on its notches, we
attached the stick to a subwoofer and vibrated it by different frequencies and
amplitudes (Fig. \ref{apparatus}).

We also changed the position of the stick on the
subwoofer (from center towards the end) to simulate different combinations of
horizontal and vertical vibrations. The phenomenon was recorded by a high
speed camera that allowed to track both the movement of the end of the stick
and the propeller.

This configuration has several advantages. First, both the amplitude and the
frequency of the oscillations are well controllable. The frequency is uniquely
defined by the frequency of the sound from the signal generator; amplitude of
oscillations depends on more parameters - amplitude of the signal, its
frequency and exact positioning of the device on the subwoofer, but is
perfectly stable once these parameters are fixed. Importantly, the movement of
the stick is harmonic in both directions, contrary to the intrinsically
inharmonic movement resulting from collisions between the sticks. This all
allowed us to perform well controllable experiments in order to distill the
leading physical effects that influence the motion of the propeller.

\subsection{Experimental conditions}

To keep the experimental setup as close to the original problem as possible,
let us analyze the Gee-haw whammy diddle frequencies. While rubbing the stick
by hand, the hand usually makes a few (say up to three) full cycles (there and
back) on the stick per second. There are usually in the order of 10 notches on
the stick which results into say up to $60$ impulses per second. This is
however only a very rough estimate, as the hand does not move with a constant
velocity there and back (resulting in a continuous range of frequencies) and
the inharmonic impulses themselves produce a broad Fourier spectrum in
oscillations. Using a subwoofer with frequency range from $20$ Hz to $90$ Hz
allowed us to simulate the leading frequencies which, as shown below, turn to
be enough to reproduce the effect -- $30$ Hz was chosen for experiments with
frequency taken as a fixed parameter. This itself is an important observation,
as some of the hypotheses \cite{Wilson, Bhattacharjee} suggested a
synchronization effect between the oscillations of the stick and the rotation
of the propeller, which would require a broad spectrum of frequencies (or
their time dependance) for starting the rotation.

The amplitudes of the motion also mimic the original problem. While on the
lower end there was no technical limitation, the maximal amplitudes were
restricted by the construction of the subwoofer. For central placement of the
stick and optimal frequencies of the subwoofer the vertical amplitude reached
as much as $10$ mm, where for edging frequencies and border placement it was
closer to a few mm. This again turned to be sufficient as we were able to
reach the optimal conditions for propeller rotation well within the range.

\begin{figure}
\begin{center}
\includegraphics[width = 0.5\linewidth]{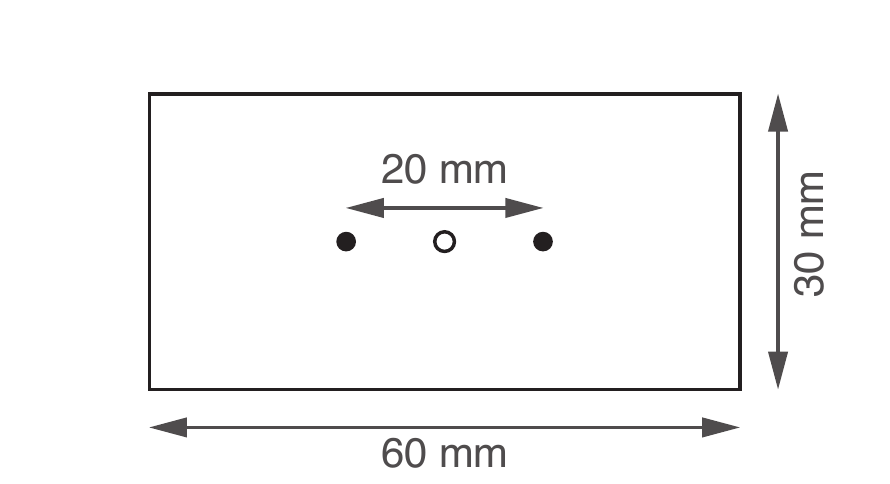}
\caption{A $60$ mm $\times$ $30$ mm propeller with a $2$ mm diameter hole.}
\label{Fig_propeller}
\end{center}
\end{figure}

It turned out that while the construction of the rod (a simple stick with a
nail on its end) does not crucially influences the movement, this is not the
case for the propeller (so much neglected in the previous work). This is
mainly due to the fact that if the center of gravity does not coincide with
the center of the hole in the propeller, the propeller has a preferred
position hanging on the notch and the start of the rotation is more
complicated. Even more importantly for small holes, if the center of gravity
is outside of the hole, other mechanisms then those described below are needed
to start the oscillations. Due to this we decided to use laser-cut propellers
in the shape of a rectangle with dimensions of $60$ mm $\times$ $30$ mm $\times$ $2$ mm and a hole of
different diameters ($2-10$ mm) cut in the center with great precision, with
weight of $4.1\pm0.1$ g depending on the diameter of the hole. Two points were
also engraved on the propeller $20$ mm apart (see Fig. \ref{Fig_propeller}) to allow tracking of
the exact position of the propeller.

The whole motion was recorded on a camera with optically stabilized lens
system with $1000$ fps and Full HD ($1920 \times 1090$) resolution. Position of the
nail and the points on the propeller was tracked with a tracking software
(Tracker). This allowed us to track more than $10$ points per
period even for the highest oscillations speeds (and more than $33$ for the
most extensively used frequency) with a precision in the order of a few $\upmu
$m. This is below the resolution of a single pixel ($37 ~ \upmu$m per pixel) due
to the fact that the tracking software analyzed whole area of pixels and was
able to determine the center of the area with higher accuracy. This all lead
to a well precise description of the movement of the nail.

\subsection{Data processing}

Two points on the side of the propeller were used to determine two parameters
-- position of the center of the propeller and its angle (determining the
angular velocity, the final output measure). Three complications were
connected with processing of the data.

First, two engraved points were designed to be symmetrically displaced from
the center of the propeller. Due to the fact that the engraving was performed
by a different device than the cut itself, imperfections in the order of $0.5$mm arose. This was however easily corrected by analyzing a set of frames in
the tracker software and adding a suitable offset to the position in the
middle of these two points.

The second problem was connected with the finite readout time of the camera.
Readout of the whole image took about $600 ~ \upmu $s, making it roughly
$t_{dr}=600$ns per row. This means that in the vertical position of the
propeller the upper point was read out about $200 ~ \upmu $s before the nail and
the lower point was delayed by about the same time. This naturally biases both
the position of the center of the propeller and its angle. This bias can be
removed recurrently in the following way.

Tilt of the propeller $\phi^{0}$ is calculated for each frame from raw data
without any correction. From this data the speed of the propeller $\omega^{0}$
can be determined in $0$th order. The time margin (or delay) of the point on
the propeller can be calculated by
\begin{equation}
t_{d}^{0}=\left(  y_{d}^{0}-y_{n}^{0}\right)  t_{dr}, \label{td0}%
\end{equation}
where $y_{d}^{0}$ and $y_{n}^{0}$ are the raw positions of the dot and nail
(in pixels) respectively. Using this time delay and the estimation of the
speed of the propeller $\omega^{0}$ one can calculate the first correction of
the point on the propeller that it is expected to reach in the time when the
point of the nail was recorded%
\begin{align}
x_{d}^{1}  &  =x_{d}^{0}-R\omega^{0}t_{d}^{0}\sin\phi^{0}\label{xd1}\\
y_{d}^{1}  &  =y_{d}^{0}+R\omega^{0}t_{d}^{0}\cos\phi^{0}, \label{yd1}%
\end{align}
where $R$ is the distance between to point and the nail, which can be well
approximated as the radius of the propeller (taken as half of the distance
between the two engraved points). These two points can serve as the input for
the next round of the recursive correction by obtaining a more precise tilt
and speed of the propeller. However it turned out that the first level is
sufficient enough to correct the influence of this effect below other
experimental imperfections.

There is another problem connected with the processing of the tilt of the
propeller. As it will be explained below, in many cases the motion of the
propeller is close to a free movement interrupted with short collision with
the nail. In certain cases the collision happens between the readouts of the
first and the second point on the propeller. Then the tilt in the specific
frame is wrongly calculated and this causes jumps in the calculated speed of
the propeller. There are two possible ways to tackle this problem. One,
obvious, is to simply ignore those readouts where the speed of the propeller
is significantly deviating in two subsequent points. The other one is to
detect the times of the collisions from changes in the translational movement
of the center of the propeller and ignore readouts for these frames. 
However, both these processes basically just remove part of the data gathered. 
In the results presented below we plot frequency of the propeller without any of these corrections, 
which makes it rather volatile, oscillating around its stable value, but clearly showing the 
level of readout errors. If some of the data is removed, the frequencies are just smoothed 
around its local stable values. 

\section{Observations}

\begin{figure}
\begin{center}
\includegraphics[width = \linewidth]{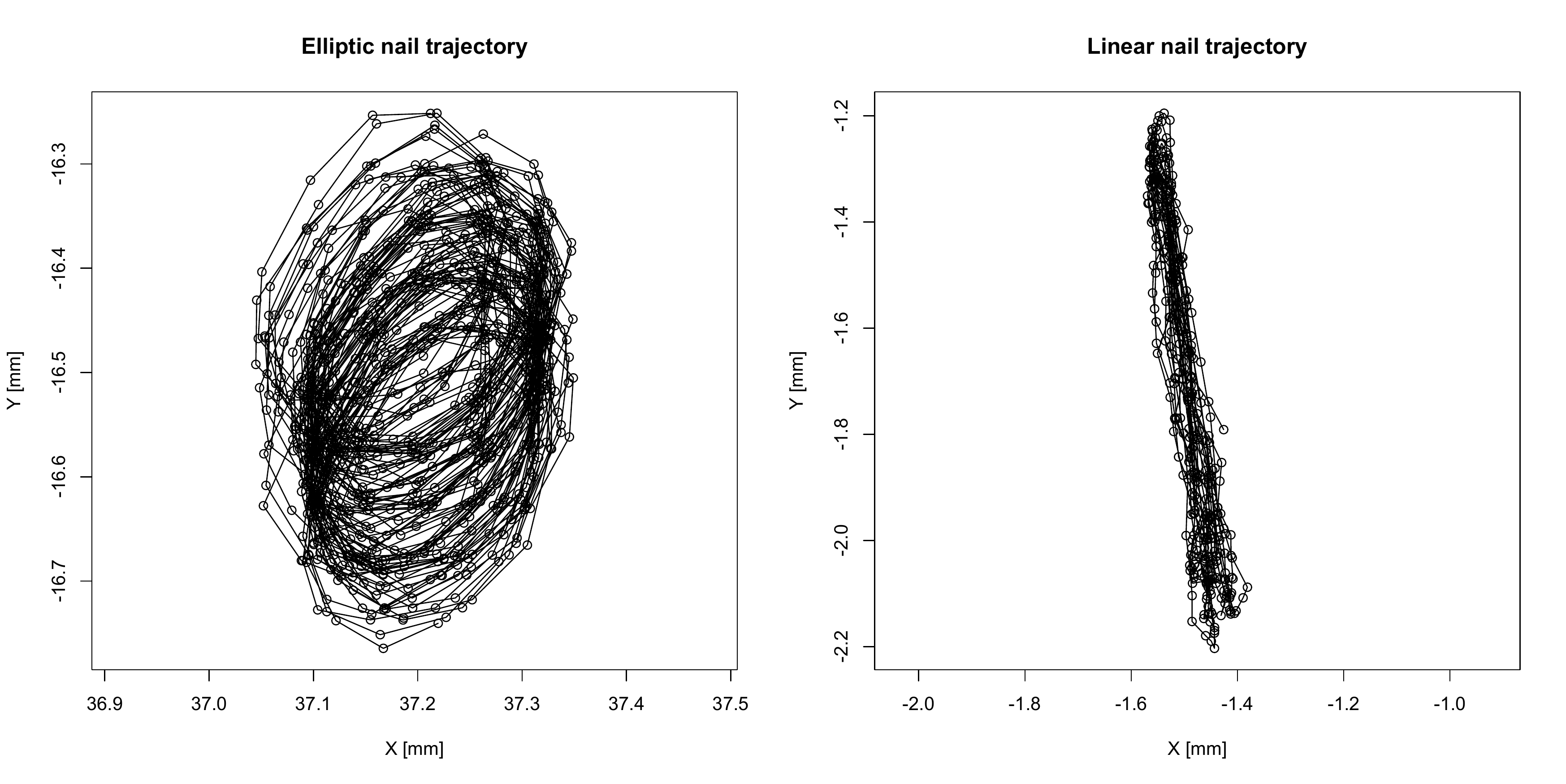}
\caption{On the left-hand side the position of the nail is depicted for the
case where the stick was placed on the side of the subwoofer. Here we can
clearly see that the movement has elliptical character with a well defined
areal velocity. On the right-hand side movement of the nail is depicted when
the stick was placed in the center of the subwoofer, where the areal velocity
is close to zero.}
\label{Fig_nail}
\end{center}
\end{figure}

We start the presentation of our results by general observations of the
movement of the propeller depending on relevant parameters. In Fig. (\ref{Fig_nail}) the readings of the position of the nail are presented for two cases -- one is elliptical, resulting from positioning of the stick on side of the subwoofer and one straight and nearly vertical, when the stick was placed in the center.
This shows that with our experimental device we were able to mimic the
outcomes produced in the close-to-original gee-haw whammy diddle in Refs.
\cite{Wilson, Satonobu}. For elliptical movement we confirmed the results
obtained so far that the direction of the rotation of the propeller was
predefined by the direction of the elliptical movement (areal velocity).
However, very interestingly, we were able to observe significant speeds of the
propeller also in the case where no areal velocity was present in the movement
of the nail (right-hand side of the Fig. \ref{Fig_nail}), although in this
case the direction of the movement was random when started from a still
position and not changed if an initial push was given to the propeller.
Therefore we concluded the proposition:

\begin{proposition}
Elliptical movement of the nail is not necessary for starting and maintaining
the rotation of the propeller and the value of the speed of the propeller is
not connected with the areal velocity of the movement of the nail.
\end{proposition}

In what follows we concentrated our effort on the purely linear movement of
the nail, opening a very interesting fundamental question -- what is the
principle that allows such a simple movement to cause the propeller to rotate?
The sole two remaining parameters of the nail movement are its frequency and
amplitude. We performed an extensive set of measurements with different
combinations of these two parameters and were measuring the speed of the
propeller without taking into account its direction. We introduce a joint
parameter of the nail movement, namely its maximal acceleration
\[
a_{n}=4\pi^{2}A_{n}f_{n},
\]
where $A_{n}$ is the amplitude of the nail's movement and $f_{n}$ its
frequency. All results obtained are combined in Fig. (\ref{Fig_acc}). Based on these results we formulate the following proposition:

\begin{figure}
\begin{center}
\input{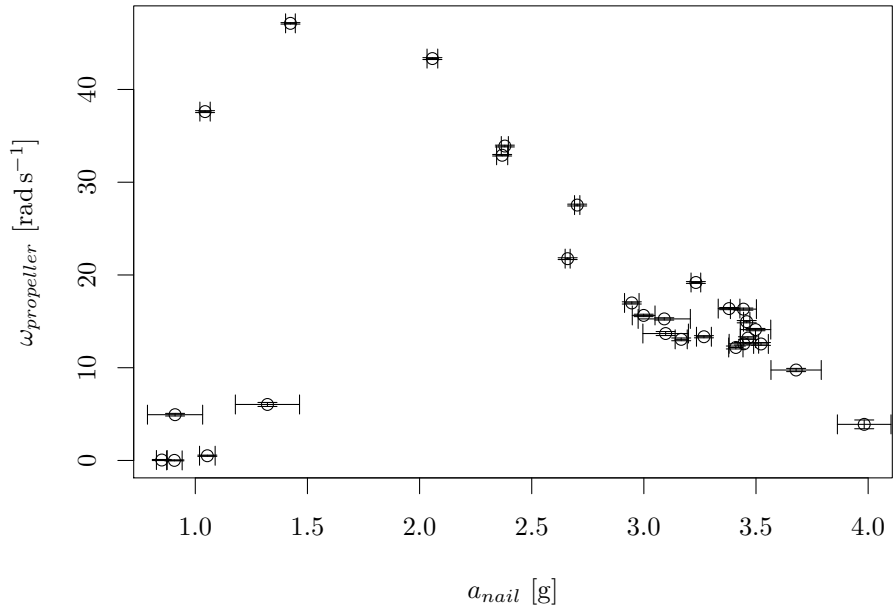}
\caption{Dependance of the propeller speed on the maximal acceleration of the
nail (measured in factors of the gravity constant $g$) during its movement.
The propeller would not rotate if this acceleration is smaller than one, i.e.
if the propeller would not de-touch from the nail even while going down. By
further increasing the acceleration, the speed of the propeller is decreasing.}%
\label{Fig_acc}
\end{center}
\end{figure}

\begin{proposition}
The propeller only rotates if the nail during its movement de-touches from the
propeller, thus moves down with acceleration exceeding gravity. Further
increase of the acceleration leads to smaller propeller speeds.
\end{proposition}

This proposition can be explained by a model where the movement of the
propeller is caused by collisions between the nail and the propeller rather
then by continuous movement of the nail along with the propeller (similar to
the way how hula-hoop is being explained). In this model, the nail needs to
de-touch from the propeller while moving down to allow an impact when
returning back. This model qualitatively also explains why the optimal speed
is reached for accelerations between $g$ and $2g$. First it is important to
notice that the mean absolute value of the acceleration is $\frac{1}{2}a_{n}$,
so if the maximal acceleration is $2g$, the "typical one" is $g$. If the
typical acceleration is around $g$, the time the propeller stays de-touched
from the nail during its way down is maximized. For higher values of
acceleration the nail quickly reaches the bottom edge of the hole in the
propeller and the driving cannot proceed. For higher acceleration also the
forces during the "touching phase" are larger and cause higher deceleration of
the propeller speed due to friction.

\begin{figure}
\begin{center}
\input{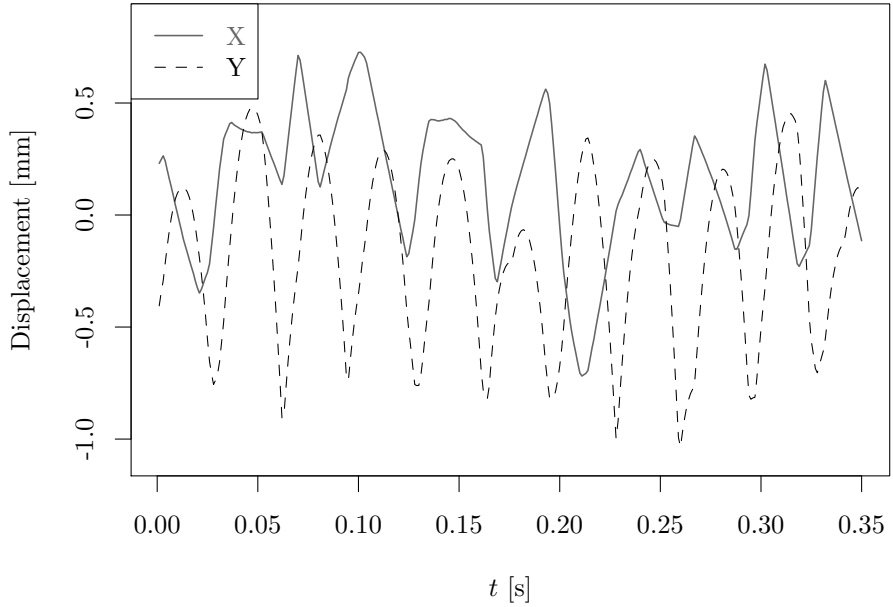}%
\caption{Here we display the position of the center of the propeller on time.
It can be seen that the x-coordinate evolves closely to free motion,
interrupted by collisions. In the y-coordinate the movement resembles
parabolas (free fall motion), again interrupted by collisions.}%
\label{Fig_propeller_center}%
\end{center}
\end{figure}

To examine this model in a more detail we focused on the movement of the
propeller. As its tilt and rotational speed are biased by fluctuations
explained above, we first concentrate on the translational movement. In Fig. (\ref{Fig_propeller_center}) we present the motion of the center of the propeller on time. It can be seen
that the motion during most of the time is governed by the free fall equation
(constant speed in x-direction and constant acceleration in y-direction),
interrupted by occasional collisions and short periods of continuous touching.

Therefore we formulate the proposition:

\begin{proposition}
For largest speeds of the propeller, its motion can be described by free fall
motion and short collisions with the nail.
\end{proposition}

\section{Model}

Based on the propositions formulated in the previous section we built up a
theoretical model. The model is based on the assumption that the movement of
the propeller is driven mainly by short collisions between the propeller and
the nail and most of the time moves freely. This assumption is backed up by
the observation of the translational movement depicted in Fig.
(\ref{Fig_propeller_center}), but also on the direct observation of video
recordings from which it is clear that most of the time the propeller does not
touch the nail.

Naturally, the crucial point in this model is to describe the collisions, as
(up to air friction) the movement in between the collisions is trivial.
Experimental data allows as to describe with a high precision the position of
the center of the nail and, knowing its diameter, also its boundary. We can
also describe the center of the propeller and boundary of the hole, but due to
imperfections based on the readout delay, exactly in those frames in which a
collision occurred this description is inaccurate. So basically for most of
the frames we get perfect data, that allows us to pinpoint both the nail and
the propeller, but for the interesting frames including collisions this is not
the case.

Due to this we had to build up a model that does not relay on detecting and
describing the collisions based on comparing the positions of the nail and the
propeller, but rather on abrupt change of translational velocity of the
propeller. For a very short collision that happens at an angle $\phi$ between
a vertically moving nail and the propeller we model the impact by two integrated forces:
\begin{equation}
 F_{perp} = \int Fdt
\end{equation}
acting on the propeller outwards from its center 
\begin{equation}
 F_{friction} = f \int Fdt
\end{equation}
in perpendicular direction, where $f$ is the friction coefficient between the
nail and the propeller, which can be negative -- the sign determines in which
direction the friction force is acting. In principle the friction force can be
smaller as well if there is no slipping on the touching
point, which is however not possible for fast moving propellers, so we do not
take this possibility into account.

These two impacts will change both the momentum in $x$ and $y$ direction of
the propeller, as well as the angular momentum of the propeller in the
following way:%
\begin{align}
\Delta p_{x}  &  =\cos\phi\int Fdt+\sin\phi f\int Fdt\label{delta_px}\\
\Delta p_{y}  &  =\sin\phi\int Fdt-\cos\phi f\int Fdt\label{delta_py}\\
\Delta L  &  =rf\int Fdt, \label{delta_L}%
\end{align}
where $r$ is the radius of the hole in the propeller. There is no experimental
way how to directly measure $\int Fdt$ for an individual collision, but we can
treat it as a variable and solve for it from the first two equations, and plug
it into the last one. We get
\begin{equation}
\Delta L=r\left(  \Delta p_{x}\sin\phi-\Delta p_{y}\cos\phi\right)  ,
\label{delta_L_res}%
\end{equation}
which can be directly calculated for each frame by knowing the $\Delta p_{x}$,
$\Delta p_{y}$ and $\phi$. Change of momentum in x direction in the collision
is simply given by its change between the frames, as there are no other forces
acting in this direction. This is not the case for $\Delta p_{y}$, where one
need to subtract the contribution of the gravitational acceleration given by
\begin{equation}
\Delta p_{y}^{g}=mgt_{f}, \label{delta_p_grav}%
\end{equation}
where $m$ is the mass of the propeller and $t_{f}=1ms$ is the time of one
frame. The angle $\phi$ can be also determined from the position of the nail
and the propeller. It is interesting to see that the resulting change in
angular momentum (\ref{delta_L_res}) is independent on the friction
coefficient $f.$

Let us recall here that the sum of the momenta in $x$ direction
(\ref{delta_px}), as well as in $y$ direction (\ref{delta_py}) corrected for
the gravitational force (\ref{delta_p_grav}) shall be close to zero for any
subsequent set of frames, which can serve as a sanity check on the data. On
the other hand if the propeller accelerates its rotation or even keeps it
constant, the sum of angular momentum contribution should be positive or
negative (depending on the direction of the rotation) and increasing its value
for an increasing number of subsequent frames.

\section{Results}

\begin{figure}
\begin{center}
\input{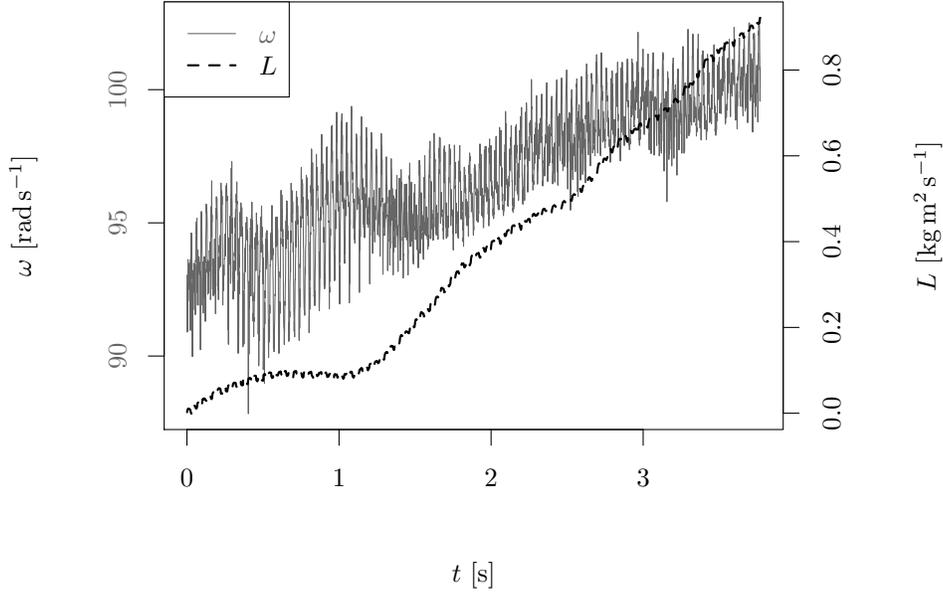}%
\caption{Here we depict the time dependence of the propeller rotation speed
$\omega$ and the aggregated angular momentum $\Delta L$. While the precision
of the propeller frequency is not very high, the increasing angular momentum
is clearly visible. Recall that part of this momentum is continuously lost in
friction.}%
\label{positive_omega}%
\end{center}
\end{figure}

\begin{figure}
\begin{center}
\input{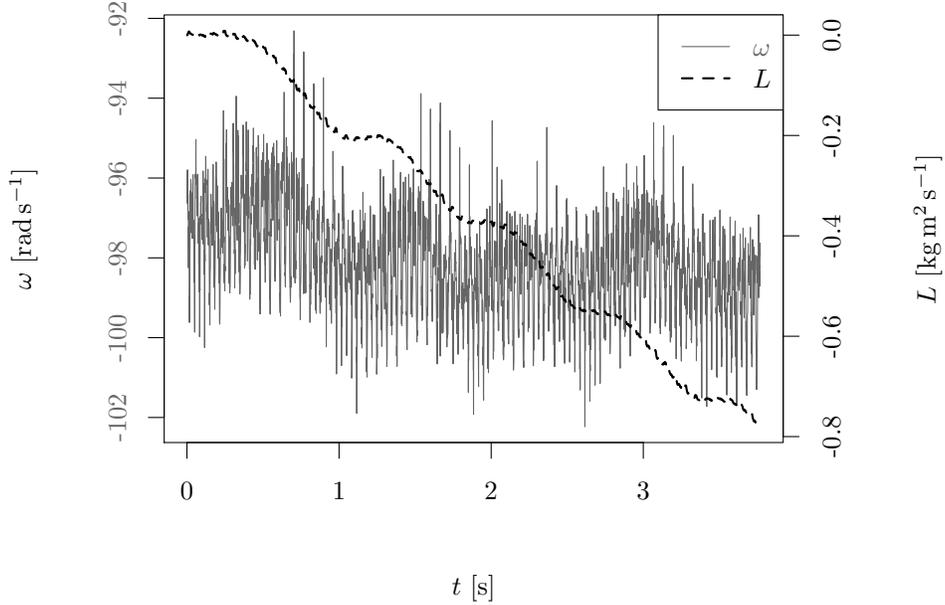}%
\caption{Here we depict the same data as in Fig. (\ref{positive_omega}) for the opposite direction
movement. It can be seen that both the propeller speed and the aggregated
angular momentum through equals time spans are basically the same. }%
\label{negative_omega}%
\end{center}
\end{figure}

We have analyzed the angular momentum contributions (\ref{delta_L_res}) for
different types of motion of the propeller -- running rotation in both
directions and rotation accelerating from still point. In Fig. (\ref{positive_omega}) we can see that the aggregated angular momentum is clearly increasing with
time. Part of this momentum is lost in friction and part is utilized for
increasing the frequency of the propeller. The same kind of data is plotted in Fig. (\ref{negative_omega}) for the negative direction of propeller movement. Despite the direction, all
other results are very similar: both the final frequency and the aggregated
angular momentum per second.

\begin{figure}
\begin{center}
\input{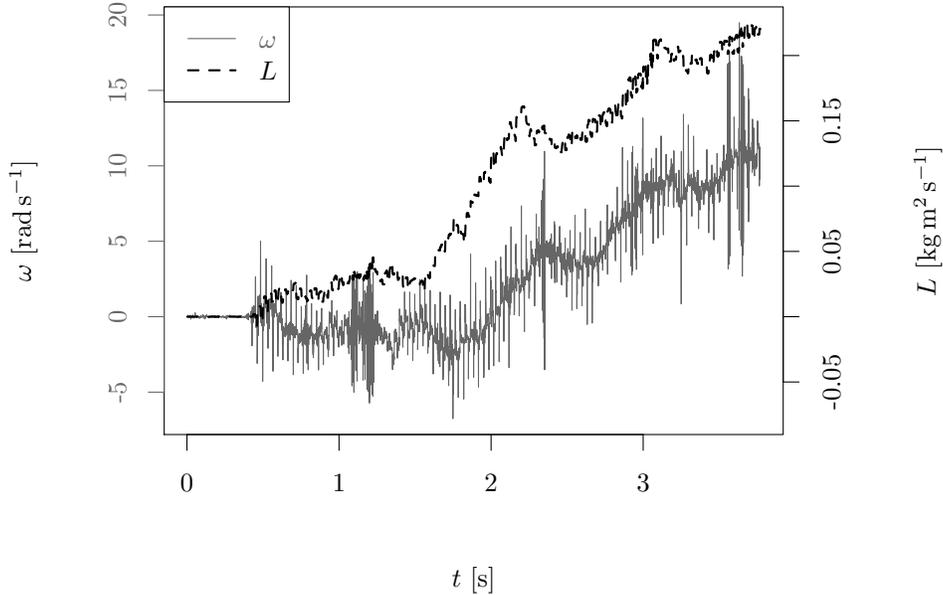}%
\caption{Starting phase of the propellers motion. It can be clearly seen that
especially in the first phase (first second) the aggregation of angular
momentum is very small and the propeller does not move. But also during the
next two seconds the efficiency of the aggregation is smaller compared to the
situation where maximal velocity is reached. }%
\label{start}%
\end{center}
\end{figure}

We also analyzed the starting period of the movement in Fig. (\ref{start}).

Here less of the aggregated momentum is lost in friction (so the propeller
accelerates), but on the other hand the efficiency of the aggregation is much
smaller, especially in the starting phase.

In principle, one can divide the frames of the recording into two types, with
and without collisions. In the first one $\Delta p_{x}=0$ and $\Delta
p_{y}-\Delta p_{y}^{g}=0$ (\ref{delta_p_grav}), leading to $\Delta L=0$,
whereas in the others these values should be nonzero (and, for relevant
collisions, not to small). However, taking the experimental errors into
account, in all frames both $\Delta p_{x}$ and $\Delta p_{y}-\Delta p_{y}^{g}$
are non-zero. One would expect that the influence of these frames is
negligible for the total aggregated angular momentum. To test this hypothesis
we can set a threshold for the total change in the momentum
\begin{equation}
\Delta p_{tot}=\sqrt{\Delta p_{x}^{2}+\Delta p_{y}^{2}}\label{total_momentum}%
\end{equation}
per frame and count only aggregated momentum from frames for which this
treshold is reached. The other equivalent option is to take a specific number
of frames with the highest change of translational momentum
(\ref{total_momentum}) and look at the total aggregated angular momentum.

\begin{figure}
\begin{center}
\input{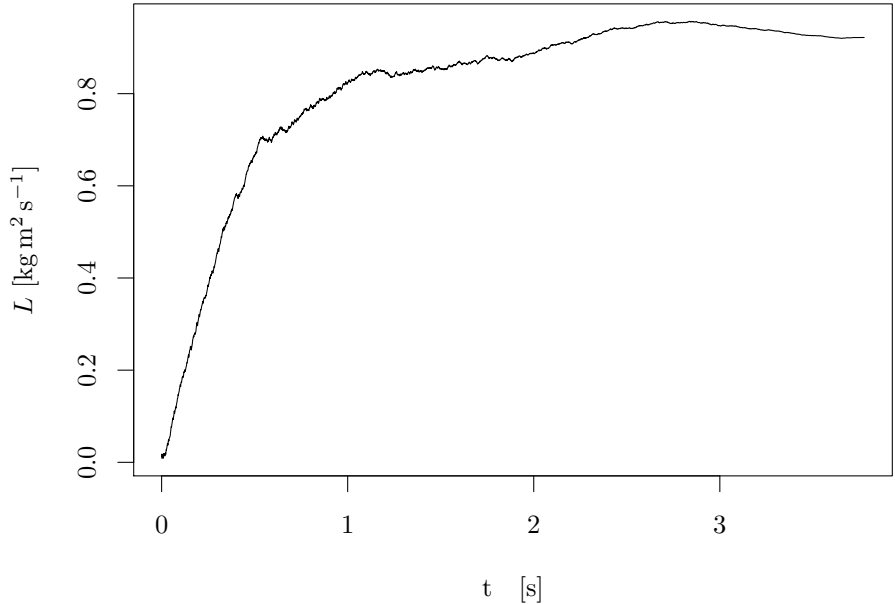}
\caption{Here we show the aggregation of angular momentum for the positive
direction of rotation based on the frames with highest change of translational
momentum, corresponding to strongest collisions. One can see that most of the
angular momentum is aggregated in about $10\%$ frames with highest change in
translational momentum, which corresponds to the observation that in about
this fraction of frames a collision occurs. }
\label{selected_frames}
\end{center}
\end{figure}

In Fig. (\ref{selected_frames}) we show the dependance of the total aggregated angular
momentum depending on the number of frames taken. Is can be clearly seen that
the main aggregation happens in about $15\%$ of the frames with highest change
of translational momentum, corresponding to the collisions. This corresponds
also to the observation of the video recordings that a collision happens a few
times during one period of the nail movement. Contribution of the rest of the
frames composes from the contribution of weaker collisions and noise and in
total is insignificant. This confirms the proposition that the propeller gains
most of the angular momentum causing rotation within short collisions rather
than by continuous touching of the nail.

\section{Conclusion}

Gee-haw whammy diddle is a very interesting mechanical scientific toy that
demonstrates the changes of linear driving to rotational movement. Within this
paper we have performed an extensive experimental research of the devise,
aiming to pinpoint the essential phenomenon that causes the propeller to
rotate. Contrary to hypotheses presented in previous works on this topic, in
particular in \cite{Wilson, Satonobu}, we have shown that the elliptical
movement of the stick is not necessary to induce the rotational movement of
the propeller -- on the contrary, purely linear motion of the stick induces the
highest speeds of the propeller. We have also shown that in contrary to ideas
presented in \cite{Wilson, Bhattacharjee} there is no synchronization effect
between the motion of the tick and the propeller -- by purely harmonic motion
of the stick the propeller was able to accelerate from rest to high
frequencies (however not reaching the driving frequency) and the stabilized
speed of the propeller was strongly dependent on the amplitude of the
oscillations.

We showed that the highest speed of the propeller can be reached for
parameters of the oscillations that secure the typical acceleration of the
stick to be roughly equal to gravitational acceleration. This is due to the
fact that, as concluded from the motion of the propeller and direct
observation, best results are achieved for situations where most of the time
the propeller is not touching the nail at the end of the stick. In this case
the propeller is being translationally stabilized by short collisions with the
moving nail that, in total, supply a flow of angular momentum. The direction of
rotation is, for linear stick movement, random.
Gee-haw whammy diddle is often compared to hula-hoop, also a well-known effect
used for physics demonstration. We believe that this comparison is rather
imprecise due to the fact that the governing physical effects are pretty much
different in those two cases. One could probably compare gee-haw whammy diddle
better to the Devil stick juggling, where one stick is being held in hand and
used to rotate another stick freely flying in the air.

\section*{Acknowledgments}

%%%%%%%%%%%%%%%%%%%%%%%%%%%%%%%%%%%%%%%%%%%%%%%
Work was motivated by the IYPT 2017 problem number 14 (www.iypt.org). This
research was supported by the VEGA 2/0043/15 project. MP would like to thank
PosAm for generous support.

\end{document}